# FLSH - Friendly Library for the Simulation of Humans

Pablo Ramón, Cristian Romero, Javier Tapia, Miguel A. Otaduy
Universidad Rey Juan Carlos, Madrid, Spain

https://gitlab.com/PabloRamonPrieto/flsh

**Abstract**

Computer models of humans are ubiquitous throughout computer animation and computer vision. However, these models rarely represent the dynamics of human motion, as this requires adding a complex layer that solves body motion in response to external interactions and according to the laws of physics. FLSH is a library that facilitates this task for researchers and developers who are not interested in the nuisances of physics simulation, but want to easily integrate dynamic humans in their applications. FLSH provides easy access to three flavors of body physics, with different features and computational complexity: skeletal dynamics, full soft-tissue dynamics, and reduced-order modeling of soft-tissue dynamics. In all three cases, the simulation models are built on top of the pseudo-standard SMPL parametric body model.

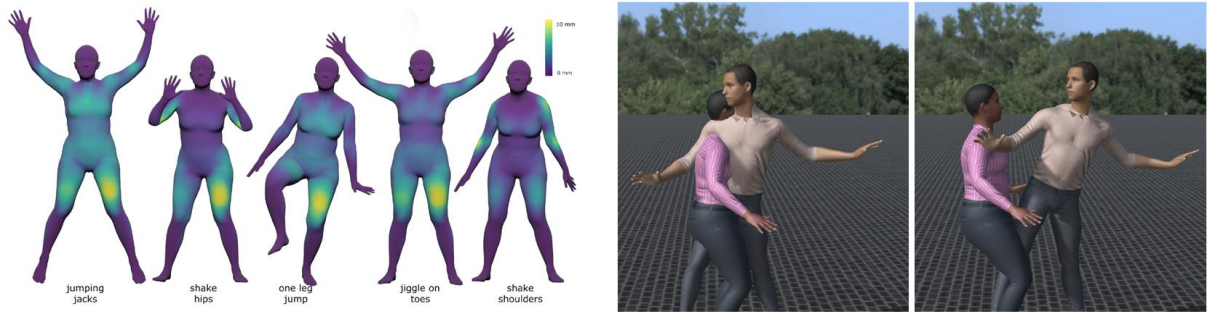

Fig. 1. Two examples that evidence the relevance of body physics. On the left, skin dynamics (color-coded) induced by skeletal motion. On the right, body-body contact ensures collision-free avatar pose and motion.

## 1. Introduction

Parametric models of humans [2] have greatly simplified the modeling, animation, reconstruction and tracking of humans across computer vision and computer graphics applications. However, to date these parametric models are intrinsically geometric, and their shape, pose and motion must be either designed or inferred through optimization methods. Parametric human models lack the ability to move according to the laws of physics. We believe that endowing parametric human models with physics-based dynamics can open up novel application possibilities, across animation, virtual reality, or computer vision.

The images in Fig.1 highlight the relevance of body physics in two examples: (left) color-coding of the amount of soft-tissue dynamics in 5 different motions; (right) extreme differences in avatar interaction without or with contact mechanics. Deformation, dynamics and contact are obtained in a natural way by solving the laws of physics on top of parametric human representations. To simplify the adoption of this technology across human modeling, animation, reconstruction and tracking, we provide FLSH, a simple software library with dynamic simulation capabilities for the popular SMPL parametric human model [2]. In FLSH, the core library runs in C++, but it offers both C++ and Python APIs to allow for flexible integration.

In FLSH, the simulation of human bodies is presented in three different flavors:

1. <u>Articulated skeleton:</u> This is an articulated skeleton dynamics model with geometric skinning. Bodies react to contact, and they can be controlled according to target trajectories or forces. This is the simplest and fastest representation, when tissue deformation due to contact or dynamics is not necessary.

2. <u>FEM soft skeleton:</u> This is a full finite-element model (FEM) of soft-tissue dynamics on top of the articulated skeleton. Bodies contain both the bone structure and deformable soft-tissue, simulated with state-of-the-art finite-element non-linear elasticity, for high-accuracy applications.
3. <u>ROM soft skeleton:</u> This is a reduced-order model (ROM) of the FEM soft skeleton. Bodies solve a real-time approximation of soft-tissue deformation, including contact, following a research solution [6]. This representation offers an optimal trade-off between accuracy and performance.

The rest of the document is structured as follows. First, we describe the different simulation models supported in the library, covering a range of features and allowing maximal performance for each feature set. Second, we outline the high-level API of the simulation library. Third, we show and discuss some simulation results.

## 2. Simulation Methods

All three simulation models in FLSH are solved using a common methodology. This simplifies the transition between simulation models based on user needs. The simulation methodology is also easy to extend. In the rest of this section, we present the overall algorithm, the main characteristics of all three simulation models, and some performance comparisons. Fig. 2 shows a comparison of avatar deformation under the three simulation models.

### 2.1. Overall algorithm

All simulation models are defined on top of the parametric body model SMPL [2]. SMPL is characterized by a pose $\theta$ and shape $\beta$. Given a template body $\bar{x}_0$ in reference pose, it is mapped to a person-specific body $\bar{x} = f_{shape}(\bar{x}_0, \beta)$ based on the shape parameters. Then, a skinning transformation $x = f_{skinning}(\bar{x}, \theta)$ based on $\theta$ defines the posed body shape $x$. SMPL's skinning transformation is linear blend skinning augmented with pose and shape blend-shapes. Even though the original SMPL is defined only for the body surface, FLSH is built on top of the extension to a volumetric body by [5].

Let us consider a set of generic dynamic degrees of freedom (DoFs) $q$, which depend on the specific simulation model of choice. The DoFs $q$ alter the standard SMPL definition of the deformed body $x$, as discussed below for each simulation model.

FLSH simulates dynamics using backward Euler numerical integration. This can be formulated as the following optimization problem [1]:

$$q = \arg\min \frac{1}{2h^2}(q - q^*)^T M (q - q^*) + V(q)$$

In this optimization, $h$ is the time step, $q^*$ the tentative state resulting from explicit integration, $M$ is the mass matrix, and $V$ collects all potential energy terms.

In essence, the API of FLSH provides the evaluation of the objective, gradient and Hessian of the optimization. This allows extensibility, as users can program their own DoFs and energy terms, which are added to those of FLSH. Furthermore, FLSH offers separate access to the inertial and potential-energy terms in the optimization. In this way, simply by dropping the inertial term, FLSH can also simulate static deformations. The library is complemented with a solver example.

FLSH offers some default contact handling capabilities, as well as tools for generalizing contact handling. First, it allows the internal definition of some simple parametric colliders (spheres, cylinders). Second, it provides access to the Jacobian $\frac{\partial x}{\partial q}$. In this way, users can define externally arbitrary energy terms $V_{users}(x)$, and obtain gradients $\frac{\partial V_{user}}{\partial q} = \frac{\partial V_{user}}{\partial x} \frac{\partial x}{\partial q}$ and approximate Hessians $\frac{\partial^2 V_{user}}{\partial q^2} = \frac{\partial x^T}{\partial q} \frac{\partial^2 V_{user}}{\partial x^2} \frac{\partial x}{\partial q}$ with respect to the DoFs.

### 2.2. Articulated skeleton

For each body bone in the SMPL skeleton, we define a rigid transformation $\phi_i$. Then, the DoFs of the skeletal dynamics model correspond to the collection of rigid bone transformations, $q = \{\phi_i\}$. We can obtain the SMPL pose $\theta$ by extracting the skeleton root transformation and the joint angles from $q$, and the deformed body $x$ is directly obtained through the skinning transformation defined above.

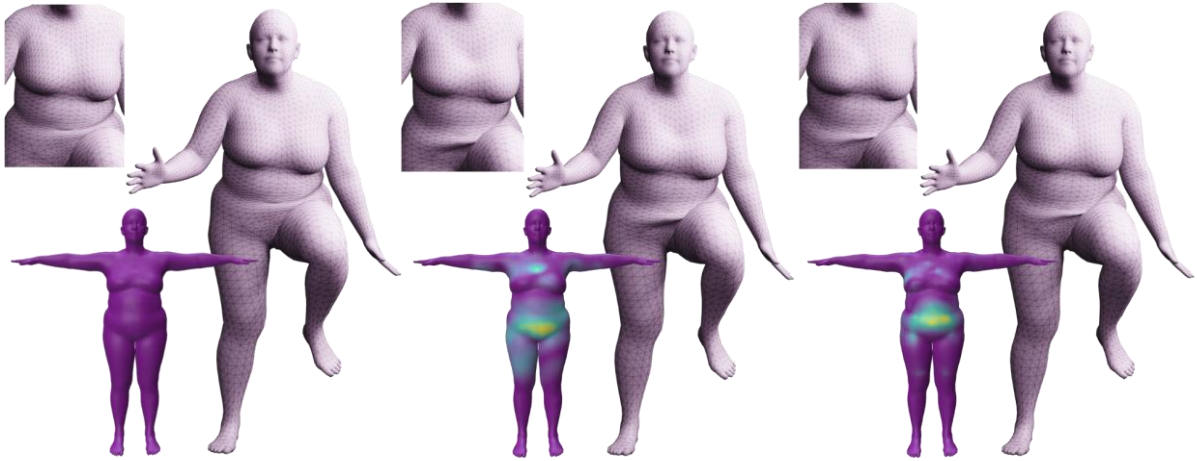

*Fig. 2. Comparison of deformations produced by the simulation modes in FLSH. From left to right: articulated skeleton, FEM soft skeleton, and ROM soft skeleton. The colormaps indicate the amount of soft tissue deformation. As shown in the images, the ROM soft skeleton provides a good approximation to the FEM soft skeleton, and therefore offers a good tradeoff between speed and accuracy.*

In the articulated skeleton model, the mass matrix $M$ is derived from rigid-body inertia terms associated with each bone. The potential energy terms included in the model are: joint constraints to maintain the skeletal structure, and tracking constraints between the skeleton and target configurations (e.g. coming from mocap) implemented as joint rotation springs.

### 2.3. FEM soft skeleton

In the FEM soft skeleton model, we define a skin displacement field $u(\bar{x})$. We apply the skin deformation in unposed state, hence transforming the reference body shape as $\bar{x} + u$. By adding SMPL's skinning transformation, the deformed body can be described as $x = f_{skinning}(\bar{x}, \theta)$. We use an FEM discretization with nodal displacements $\{u_j\}$. Grouping all nodal displacements together with the rigid bone transformations yields the degrees of freedom $q = (\{\phi_i\}, \{u_j\})$. In FLSH, all body shapes use the same discretization, independent of the shape coefficients $\beta$. This allows easy transition between body shapes if needed by the user's application.

We model skin elasticity using the approach in [5], which implies:

1. The definition of deformation gradient $F = I + \frac{\partial u}{\partial \bar{x}}$, which accounts for contact and inertial deformations, but respects SMPL's data-driven detail under pose changes.

2. A strain energy density $\Psi(F)$ corresponding to an orthotropic Saint Venant-Kirchhoff hyperelastic model with Fung-type saturation. The soft-tissue parameters are estimated from the DYNA dataset [3] following the multi-person optimization approach of [4].

In the FEM soft skeleton model, the energies of the articulated skeleton are extended in the following way. The mass matrix $M$ is augmented with terms corresponding to the skin displacements, as well as cross-terms with the bone transformations. The new terms are derived from the kinetic energy of world-space skin velocities $\dot{x}$. The potential energy $V$ now also includes the integral of the strain energy density $\Psi$ over the body's volume.

### 2.4. ROM soft skeleton

The reduced model retains much of the structure of the FEM soft skeleton, but it represents the skin displacement field in a linear reduced basis, $u(\bar{x}) = U(\bar{x})\,z$, with $z$ the vector of reduced DoFs and $U$ the matrix of basis coefficients. We use as reduced model the bounded generalized biharmonic coordinates of [7], following the model architecture of [6]. This allows seamless integration in the basis of handle points in the skin together with the skeletal bones. The DoFs of the ROM soft skeleton model are $q = (\{\phi_i\}, z)$. The deformed body $x$ is defined with the same expression as in the FEM soft skeleton, with the only difference that skin displacements are represented in the reduced basis.

Using a reduced basis instead of a full FEM discretization dramatically reduces the solver cost, thanks to the reduced size of the problem gradient and Hessian. We also add cubature, following again the data-oblivious approach of [6], to approximate the skin inertia and elastic energy.

### 2.5. Performance

The performance of the various simulation models is highly dependent on the number of degrees of freedom. The SMPL model, which serves as basis for all models, consists of 24 bones, 13,776 triangles and 6,890 vertices. Pose $\theta$ is parameterized with 144 coefficients and shape $\beta$ with 10 coefficients. Our tetrahedral volume discretization of the body for the FEM soft skeleton consists of around 43500 tetrahedra and 10400 nodes. Furthermore, the ROM skeleton model contains 87 point handles, and we use 500 cubature samples.

In our performance tests, we have measured both the runtime cost and the generation of the avatars. Note that the soft models require a more costly generation process which includes tetrahedral meshing and, in the case of the ROM soft skeleton, cubature sampling. The current version of FLSH implements avatar generation every time a new avatar is created, but future versions will support caching of avatar model data. Table 1 shows the performance measurements on a PC with a 3.4 GHz Intel Core i7-4770 CPU with 32GB of memory.

*Table 1. Benchmarking.*

|  | **Avatar generation** | **Simulation** |
|---|---|---|
| Articulated skeleton | <1 s | 45 fps |
| FEM soft skeleton | 15 s | 0.2 fps |
| ROM soft skeleton | 25 s | 20 fps |

## 3. Library API

The FLSH library contains two main objects that are accessible through its API: the `SoftAvatar` class, which stores the different data structures to represent the avatar body, and the `Simulable` class, which provides access to simulation state and optimization terms.

### 3.1. Initialization and settings

The creation of an avatar must follow these steps:

1. Call the constructor of the `SoftAvatar` class. The constructor receives as arguments an SMPL avatar object and a vector of shape coefficients.

2. Enable settings that define the type of simulation model. There are 4 different settings:

   `enableSoftTissue()` If disabled, the simulation model is the default articulated skeleton.

   `enableReducedModel()` It allows selecting FEM or ROM soft skeleton.

   `enableInertialCubature()`

   `enableElasticCubature()`

3. Call `SoftAvatar::initVolumetricData()`.

The creation of the simulation object must follow these steps:

1. Call the constructor of a `Simulable` class. The constructor receives the following arguments: a `SoftAvatar` object, the SMPL object, and possibly a sequence of mocap frames to be tracked.

2. Call `Simulable::initialize()`.

If the user wants to include multiple avatars in the scene, then this is as simple as creating multiple `SoftAvatar` objects and `Simulable` interfaces. Each `Simulable` will read/write data referred to its state size, and this state can be merged into larger vectors and/or matrices corresponding to the full scene.

## 3.2. Runtime API

High-level per-step calls:

`preUpdate`: updates the active frame of mocap data, if mocap tracking is used.

`postUpdate`: updates internal data.

State management:

`getState`: it gets the current state of the DoFs.

`updateState`: it sets a new state computed by the external solver.

`pushState`: adds the current state to a stack. This can be used e.g. in line-search solvers.

`popState`: resets the previous state from the stack.

`getStateJacobian`: it gets the Jacobian of the skeleton pose, for external skeleton-based energies.

`getNodesStateJacobian`: it gets the Jacobian of mesh nodes, for external mesh-based energies.

Optimization evaluation:

`getDynamicEnergyScalar`: get the inertial part of the optimization function.

`getPotentialEnergyScalar`: get the potential-energy part of the optimization function.

`getDynamicGradientVector`: get the inertial part of the optimization gradient.

`getPotentialGradientVector`: get the potential-energy part of the gradient optimization gradient.

`fixVectorKinematic`: it receives a full gradient and applies Dirichlet boundary conditions.

`getDynamicHessianTripletVector`: get the inertial part of the optimization Hessian.

`getPotentialHessianTripletVector`: get the potential-energy part of the optimization Hessian.

`fixMatrixKinematic`: it receives a full Hessian and applies Dirichlet boundary conditions.

## 4. Simulation Results

Next, we discuss some simulation examples that highlight the features of the avatar models in FLSH, as well as differences between the various simulation models.

Fig. 2 compares body deformations with all three simulation models while tracking a mocap sequence. The images also highlight the differences in soft-tissue deformation. As expected, the articulated skeleton's deformation is due simply to SMPL's skinning transformation, but there is no additional soft-tissue deformation. In the soft skeleton models, on the other hand, there is soft-tissue deformation induced by the body's own inertia. The ROM soft skeleton provides a good approximation to the FEM soft skeleton, and therefore offers a good tradeoff between speed and accuracy.

As mentioned earlier, in the soft skeleton models in FLSH, soft-tissue properties are adapted based on body shape. Fig. 3 shows an example of this shape-based soft-tissue parameterization, for 5 different body shapes. As expected, the differences in soft-tissue parameters produce noticeable differences in the soft-tissue deformation of the various body shapes.

Finally, we highlight the benefit of physics-based simulation to resolve contact-based interactions. Fig. 4 shows an example of articulated skeleton where the lack of physics-based simulation leads to disturbing avatar interpenetration. As shown in the images, the FLSH avatars are endowed with a hierarchy of analytical capsules for fast collision detection. Fig. 5 compares the accuracy of contact resolution with the different simulation model in FLSH. From the images, we draw three main conclusions: (1) Lack of simulation produces a very unrealistic result. (2) In the articulated skeleton model, the full reaction to contact is absorbed by the skeleton, while in the soft skeleton models part is absorbed by the skeleton and part by the soft tissue, as evidenced by the displacement of the chin. (3) The ROM soft skeleton approximates well contact for large objects like this sphere, but the FEM soft skeleton becomes necessary to resolve contact with small objects and/or sharp features.

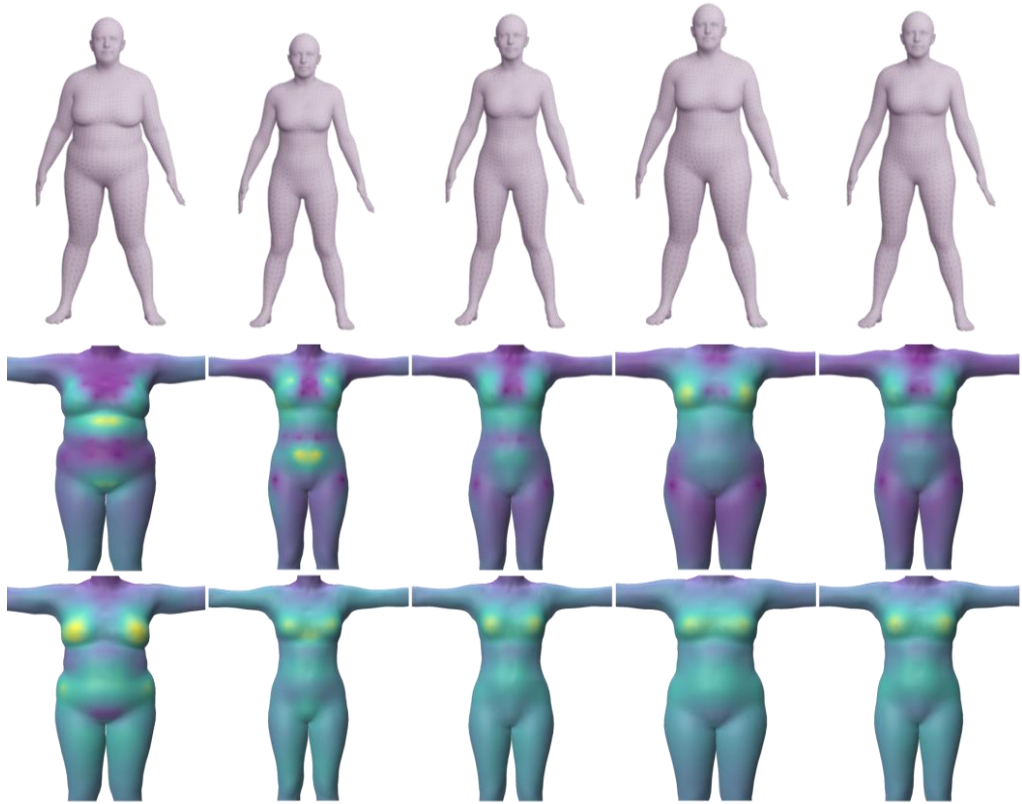

*Fig. 3. In the soft skeleton models in FLSH, soft-tissue properties are adapted based on body shape. This image shows 5 different body shapes (first row), which lead to different distribution of Young modulus (second row) as well as skin thickness (third row).*

## 5. Discussion

FLSH is conceived as a living project to offer physics-based avatar simulation to the research community. Its first version includes the three simulation models described in this paper, and we hope that adoption is facilitated by the versatile cross-platform API and the examples provided.

We understand that in many computational problems involving humans, the shape, pose or motion of the human are variables that are optimized. To provide stronger support to such applications, we would like to augment FLSH with differentiability capabilities in the future.

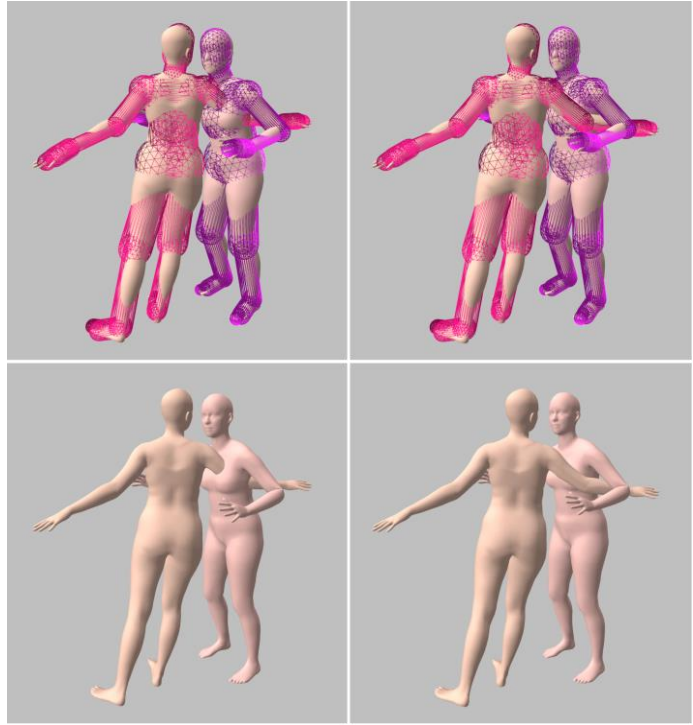

Fig. 4. Without physics-based simulation, avatars may interpenetrate each other in disturbing ways, as shown on the left. With the physics-based simulation in FLSH, on the other hand, interactions are much more accurate, as shown on the right. This example uses the articulated skeleton model. For fast collision detection, each avatar is represented with proxy geometry formed by a hierarchy of analytical capsules (shown in the top images).

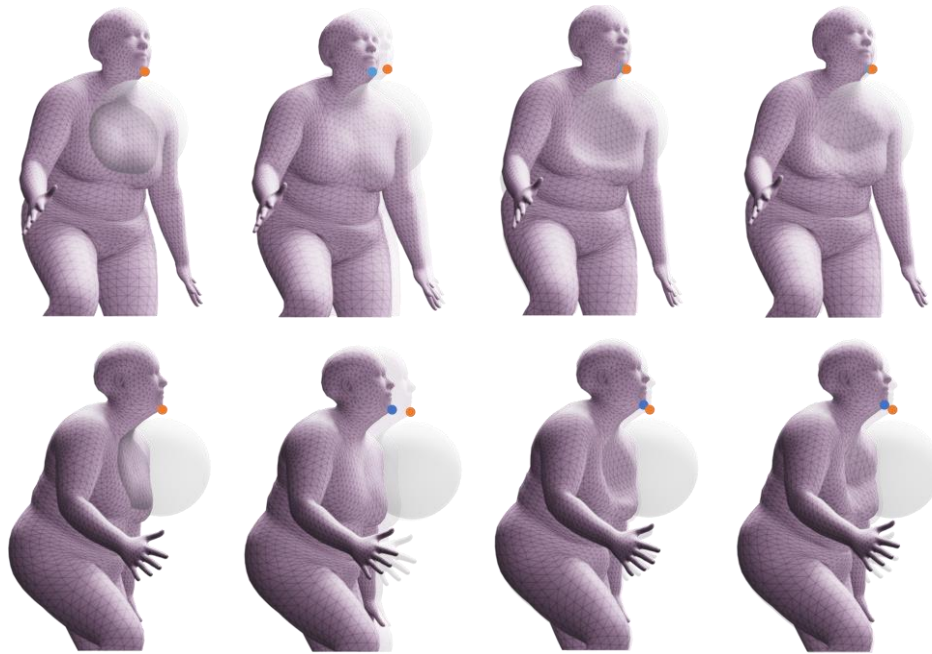

Fig. 5. Comparison of contact resolution with the different simulation model in FLSH. To highlight the differences, we display semi-transparent an avatar with no physics-based simulation, and we indicate the position of the chin in the avatars with and without simulation using colored dots. From left to right: avatar with no simulation, articulated skeleton model, FEM soft skeleton model, and ROM soft skeleton model. From the images, we draw three main conclusions: (1) Lack of simulation produces a very unrealistic result. (2) In the articulated skeleton model, the full reaction to contact is absorbed by the skeleton, while in the soft skeleton models part is absorbed by the skeleton and part by the soft tissue, as evidenced by the displacement of the chin. (3) The ROM soft skeleton approximates well contact for large objects like this sphere, but the FEM soft skeleton becomes necessary to resolve contact with small objects and/or sharp features.